\documentclass[useAMS,usegraphicx,usenatbib]{mn2e} 
\usepackage{amsmath,graphicx}
\usepackage{ulem}
\usepackage{color}
\usepackage{graphicx}
\usepackage{times} 
\usepackage{amssymb}
\usepackage{amsmath}
\usepackage{lscape}
\usepackage{url}
\usepackage{multirow}
\usepackage{hyperref}
\newif\ifAMStwofonts
\AMStwofontstrue
\newcommand{\km}{\rm\thinspace km}

\newcommand{\cm}{\rm\thinspace cm}

\newcommand{\s}{\rm\thinspace s}

\newcommand{\kmps}{\hbox{$\km\s^{-1}\,$}}

\newcommand{\pcmsq}{\hbox{$\cm^{-2}\,$}}
\newcommand{\pcmcu}{\hbox{$\cm^{-3}\,$}}

\def\h0{\mbox{{\rm H}$^0$}}
\def\he0{\mbox{{\rm He}$^0$}}

\def\H0{\mbox{{\rm H}$_{0}$}}

\begin{document}

\title[Shocks and edges in A2219] {A series of shocks and edges in Abell 2219} \author[] {\parbox[]{6.in}
  {R.E.A.~Canning$^{1,2}$\thanks{E-mail:
      rcanning@stanford.edu}, S.W.~ Allen$^{1,2,3}$, D.E.~Applegate$^{4}$, P.L.~Kelly$^{5}$, A.~von der Linden$^{6,1,2}$, A.~Mantz$^{7,8}$, E.~Million$^{9}$, R.G.~Morris$^{1,2,3}$, H.R.~Russell$^{10}$\\ } \\
  \footnotesize
  $^{1}$Kavli Institute for Particle Astrophysics and Cosmology (KIPAC), Stanford University, 452 Lomita Mall, Stanford, CA 94305-4085, USA\\
  $^{2}$Department of Physics, Stanford University, 452 Lomita Mall, Stanford, CA 94305-4085, USA\\
  $^{3}$SLAC National Accelerator Laboratory, 2575 Sand Hill Road, Menlo Park, CA 94025, USA\\
  $^{4}$Argelander-Institut f{\"u}r Astronomie, Auf dem H{\"u}gel 71, D-53121 Bonn, Germany\\
  $^{5}$Department of Astronomy, University of California, B-20 Hearst Field Annex 3411, Berkeley, CA 94720-3411, USA\\
  $^{6}$Dark Cosmology Centre, Niels Bohr Institute, University of Copenhagen Juliane Maries Vej 30, 2100 Copenhagen, Denmark\\
  $^{7}$Kavli Institute for Cosmological Physics, University of Chicago, 5640 South Ellis Avenue, Chicago, IL 60637-1433, USA\\
  $^{8}$Department of Astronomy and Astrophysics, University of Chicago, 5640 South Ellis Avenue, Chicago, IL 60637-1433, USA\\
  $^{9}$Johns Hopkins University Applied Physics Laboratory, 11100 Johns Hopkins Road, Laurel MD, 20723, USA\\
  $^{10}$Institute of Astronomy, Madingley Road, Cambridge, CB3 0HA\\
  \\}

\maketitle

\begin{abstract} 
We present deep, 170~ks, \textit{Chandra} X-ray observations of Abell 2219 ($z=0.23$) one of the hottest and most X-ray luminous clusters known, and which is experiencing a major merger event. We discover a `horseshoe' of high temperature gas surrounding the ram-pressure-stripped, bright, hot, X-ray cores. We confirm an X-ray shock front located north-west of the X-ray centroid and along the projected merger axis. We also find a second shock front to the south-east of the X-ray centroid making this only the second cluster where both the shock and reverse shock are confirmed with X-ray temperature measurements. We also present evidence for a sloshing cold front in the `remnant tail' of one of the sub-cluster cores. The cold front and north-west shock front geometrically bound the radio halo and appear to be directly influencing the radio properties of the cluster.  
\end{abstract}


\section{Introduction}
\label{intro}

\begin{figure*}
 \begin{center}
  \includegraphics[width=0.7\textwidth]{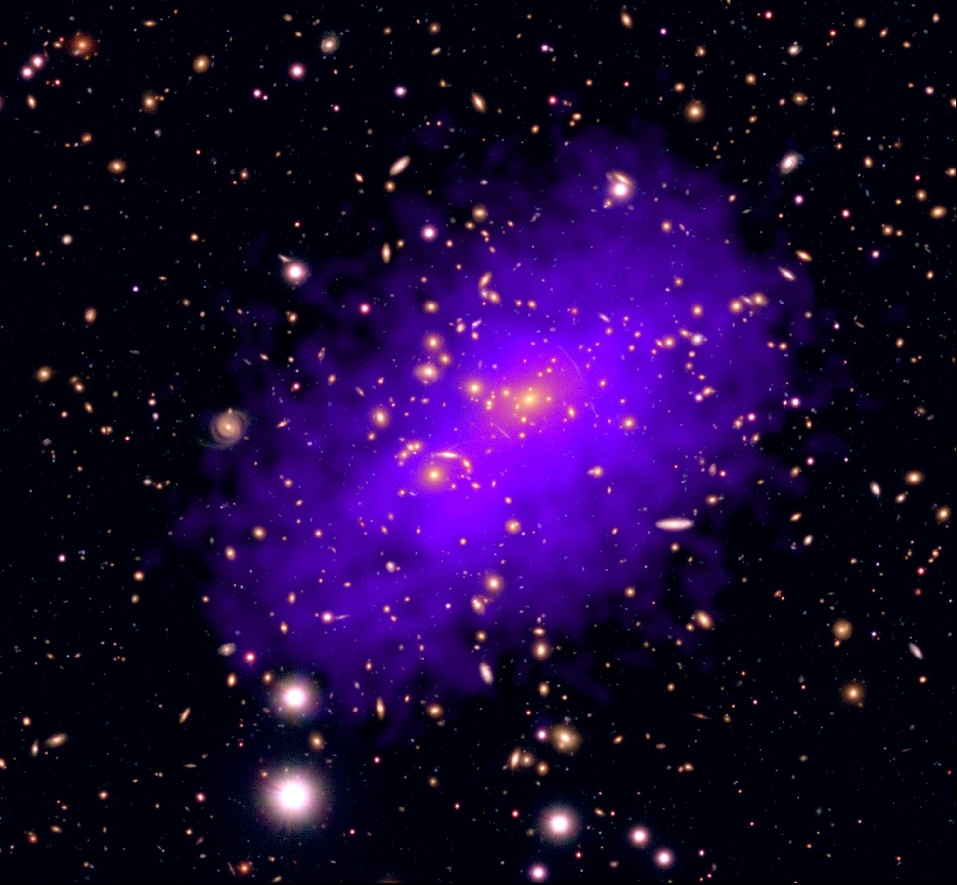}
\caption{Abell 2219: A three colour Subaru optical image using the Rc, V and B filters and the \textit{Chandra} X-ray surface brightness image (purple). \label{four_col}}
 \end{center}
\end{figure*}

Merger-driven shocks, generated during the collisions of galaxy clusters, dissipate energy into the intra-cluster medium (ICM) through the injection of turbulence and acceleration of particles, and may therefore play an important role in the evolution of large-scale structure in the Universe. Merger shocks can be observed as temperature, pressure and entropy discontinuities in the hot X-ray gas and measurements of the properties of these discontinuities can be used to constrain transport processes in the ICM (see \citealt{markevitch2007} for a review). However, observations of clear merger shock fronts, with both unambiguous temperature and density jumps, are rare. This is due to the requirements of a near plane-of-sky merger, to mitigate the effects of projection, and due to the inherently low X-ray surface brightness of shocked regions in the outskirts of clusters. To date only a handful of unambiguous X-ray shocks with clear temperature and density jumps have been discovered (e.g. the Bullet cluster, \citealt{markevitch2002}; Abell 520, \citealt{markevitch2005}; Abell 2146, \citealt{russell2010}; Abell 754, \citealt{macario2011}; Abell 2744, \citealt{owers2011}; Abell 2034, \citealt{owers2014}).

Merging galaxy clusters are often associated with large-scale (hundreds of kpc to a few Mpc), diffuse, steep-spectrum radio emission. Two flavors of this diffuse radio emission, `radio halos' and `radio relics' have been identified; the first is typically unpolarised to the few percent level and is distributed similarly to the cluster gas whilst the second is located towards the cluster periphery, often elongated in morphology and more strongly polarized $\sim$10-20 per cent (see \citealt{feretti2012} for a review). These diffuse radio components are synchrotron radiation from relativistic particles re-accelerated by turbulence or shocks (e.g. \citealt{petrosian2001, brunetti2004, brunetti2005}). They are associated with the ICM, rather than cluster galaxies, indicating non-thermal components are present on the cluster scale. Their differing morphologies suggest radio halos may be associated with turbulent reacceleration of particles in the cluster atmosphere, while the location of radio relics and their relation to surface brightness edges in X-ray observations suggest they trace shock fronts generated by major or minor mergers (see review by \citealt{bruggen2011} and references therein).

In this paper we study one such merging cluster in detail, Abell 2219 ($z=0.2256$), one of the hottest (T$_{X}=9.5$~keV, \citealt{allen1998b}) and most X-ray luminous clusters known \citep{ebeling1998}. Abell 2219 is more than twice as bright in the X-rays as either the Bullet cluster or Abell 520. X-ray and optical observations show clear evidence for a disturbed, elongated morphology and more than one dominant mass clump (\citealt{allen1998a, bezecourt2000, boschin2004, smith2005, vonderlinden2014}; the optical and X-ray morphology of the Abell 2219 system are shown in Fig. \ref{four_col}). From a multiwavelength analysis, \cite{boschin2004} suggest a complex merger history in Abell 2219, with the system undergoing an in-fall of many clumps aligned with a filament in the foreground, oriented at an oblique angle with respect to the line-of-sight. Abell 2219 also hosts a well known radio halo \citep{giovannini1999, bacchi2003, orru2007} and three strong radio galaxies \citep{owen1992}. Using the \textit{Chandra} X-ray space telescope, \cite{million2009} detected a large scale and very high-temperature shock front $\sim$2 arcmin ($\sim$430 kpc) from the cluster core. Their observed temperature and density jumps lead to Mach numbers of $>$1.3 with an estimated shock velocity of $\sim$2500~\kmps. Surprisingly, the authors found Abell 2219 also appears to contain a `hot core' ($>$20~keV). The astrophysical cause for this hot core is unclear. 

Here we present an analysis of new deep $\sim$170~ks \textit{Chandra} observations of Abell 2219. The \textit{Chandra} observations and data reduction are discussed in Section \ref{dataprep}. Results from imaging, thermodynamic maps and spectral and surface brightness profiles are presented in Sections \ref{imaging}, \ref{thermo} and \ref{profiles} respectively. We discuss the X-ray results in the context of other multi-wavelength data on Abell 2219 in Section \ref{radioxrayconnection} and \ref{discussion}, and summarize our findings in Section \ref{summary}.

We assume a standard $\Lambda$CDM cosmology with H$_{0}$ = 71 km~s$^{−1}$, $\Omega_{m}$ = 0.27 and $\Omega_{\Lambda}$ = 0.73. For this cosmology, at the redshift of Abell 2219, an angular size of 1'' corresponds to a distance of 3.588 kpc.

\section{\textit{Chandra} data preparation}
\label{dataprep}

Deep \textit{Chandra} observations of Abell 2219 were made between 2012 May 26th and 2012 October 15th. The observations were made with the Advanced CCD Imaging Spectrometer (ACIS) in VFAINT mode and using CCDs 1--4 and 6. The data were reprocessed using {\sc ciao 4.5} and {\sc caldb 4.5.9}. Existing ACIS-S data taken on 2000 March 31st was also reprocessed. The total cleaned exposure time of the observations is 169 ks. A description of the observations is given in Table \ref{description}. 

We extract and filter the lightcurves from all source-free CCDs, using energy ranges of 2.5 to 7 keV and 1024s bins for the back illuminated chips and 0.3 to 12 keV in 256s bins for front illuminated chips, respectively, also excluding point sources. These binnings match the cleaning set-up used for the blank sky background files. We use a sigma clipping algorithm to search for and reject flares. This is followed by a visual inspection of the light curves.

Blank-sky backgrounds are processed and re-projected to match the observations. The blank-sky exposures were normalised to match the particle background-dominated count rate in our corresponding target exposures, using the data in the 9.5 to 12 keV band. 

Narrow- and broad-band images between 0.6 and 7 keV were created, background subtracted and exposure corrected. Exposure map weighted PSF images were examined with the {\sc ciao} routine {\sc wavdetect} to identify point sources. A manual inspection of the point sources was then performed and these sources were excluded from the remainder of our analysis.

\begin{table}
\caption{Description of observations. OBS ID 7892 has a cleaned exposure time of less than 5 ks and as such was not included in the analysis presented in this paper. \label{description}}
 \centering
\begin{tabular}[]{c|c|c|c}
\hline
 OBS ID & Date & Instrumental & Cleaned \\
  &  & setup & exposure (s)\\
\hline \hline
 896 & 2000-03-31 & ACIS-S, 235678 & 40,250  \\
 13988 & 2012-05-26 & ACIS-I, 01236 & 8,670 \\
 14355 & 2012-06-28 & ACIS-I, 01236 & 26,100 \\
 14356 & 2012-10-15 & ACIS-I, 01236 & 41,450 \\
 14431 & 2012-05-27 & ACIS-I, 01236 & 34,700 \\
 14451 & 2012-06-26 & ACIS-I, 01236 & 17,660 \\
\end{tabular}
\end{table}

\begin{figure}
 \begin{center}
  \includegraphics[width=0.48\textwidth]{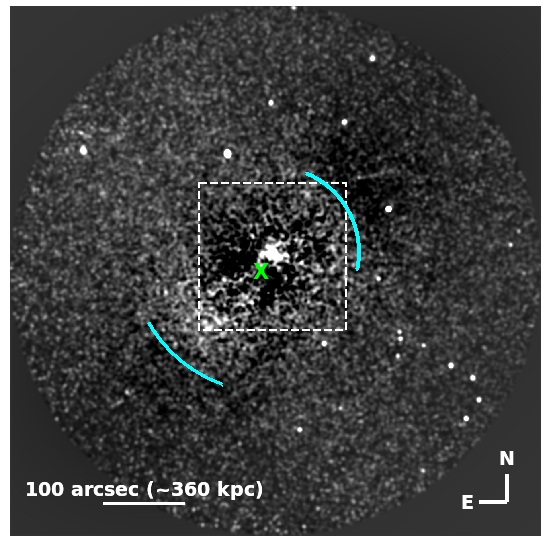}
  \includegraphics[width=0.46\textwidth]{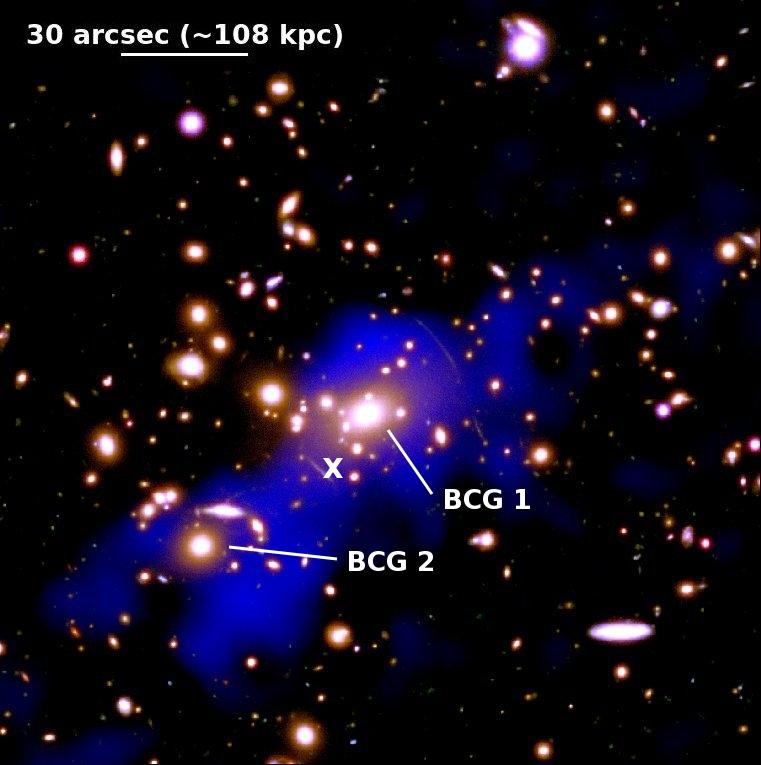}
\caption{{\bf Top:} Mosaiced, exposure corrected X-ray image, with a 2D elliptical beta model fit and subtracted. Two surface brightness edges are observed and are indicated with cyan arcs. The dashed box is 3 arcmin by 3 arcmin and is the same size as the lower panel. {\bf Bottom:} Colour composite of an unsharp-masked X-ray image (blue) (made by subtracting an image smoothed by a Gaussian with $\sigma=35$ from an image smoothed by a Gaussian with $\sigma=9$) and a three colour optical Subaru image. Two X-ray bright regions are observed in the core and are associated with the dominant galaxies, with the `peak' in the X-ray brightness associated with BCG1. The positions of the dominant galaxies and the {\it centroid} of the 2D beta model are indicated. \label{unsharpmask}}
 \end{center}
\end{figure}

We check the blank sky backgrounds against source free regions of our observations for every OBS ID. We find that on the S5 chip of the 896 ACIS-S data an additional soft component is required to model the background. Fitting an {\sc apec} \citep{smith2001} model for thermal Bremsstrahlung and line emission to this soft component, we find a best fit temperature of 1.1 keV. Overlaying ROSAT X-ray data and Digital Sky Survey optical images shows that there is an excess of both X-ray emission and a projected galaxy over-density North (N) of the S5 chip, indicating the soft excess may be emission from a group. When the region of the chip containing most of the `group' emission is excluded, no soft excess is required to fit the data.

We also examined the effect of adjusting the applied background normalisation by $\pm$5 per cent and inspect the effect on the resulting thermodynamic maps. Within their uncertainties our measurements are not affected.

\section{Imaging analysis}
\label{imaging}

The X-ray emission is elongated in the north-west (NW) to south-east (SE) direction (Fig. \ref{four_col}). Two bright concentrations of galaxies are observed along the same axis, which is the likely projected merger axis. 

\begin{figure*}
 \begin{center}
  \includegraphics[width=0.4\textwidth]{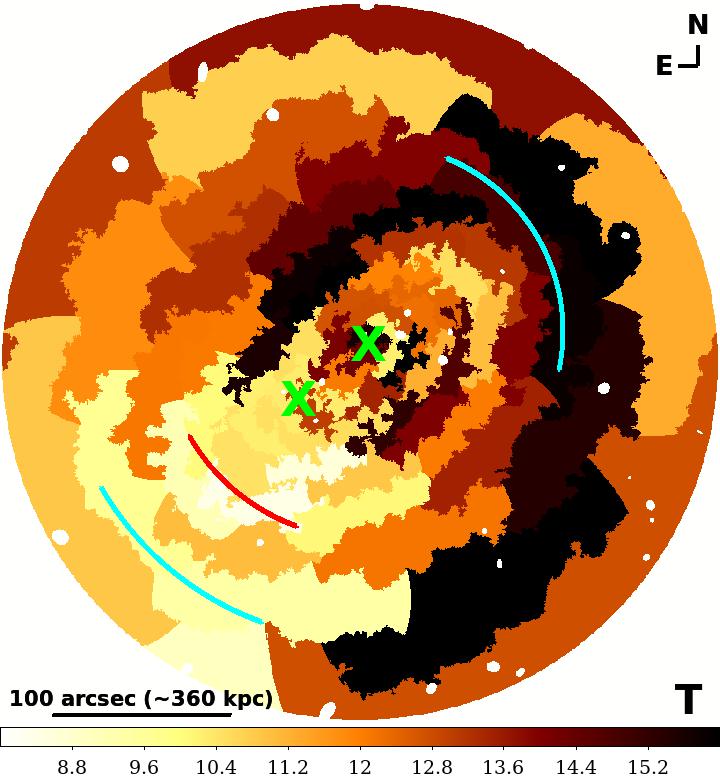}
  \includegraphics[width=0.4\textwidth]{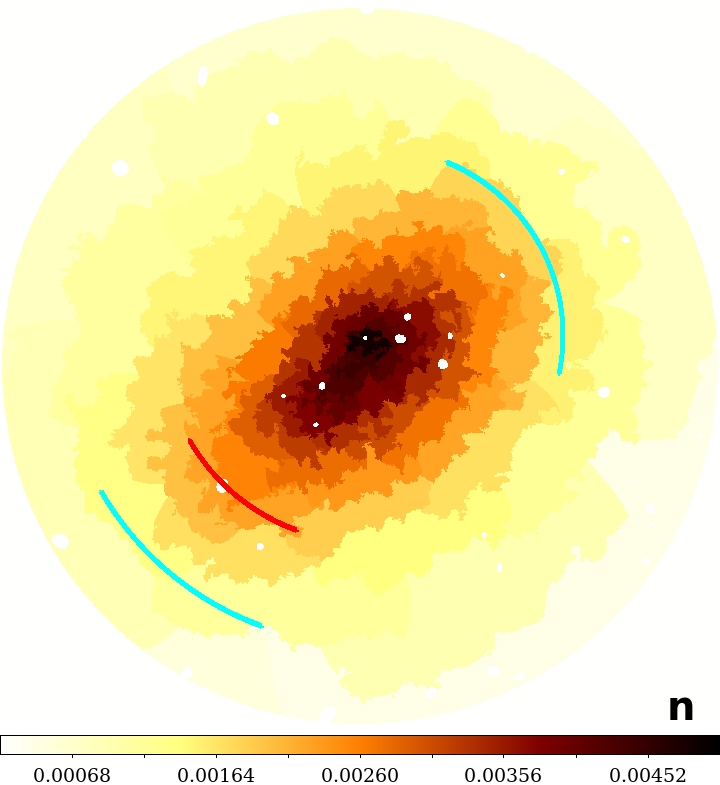}
  \includegraphics[width=0.4\textwidth]{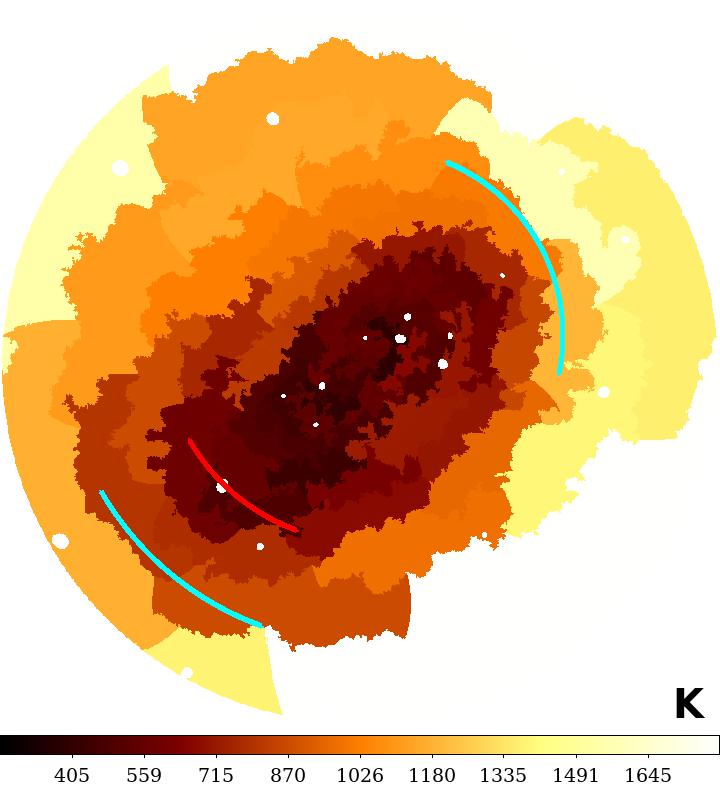}
  \includegraphics[width=0.4\textwidth]{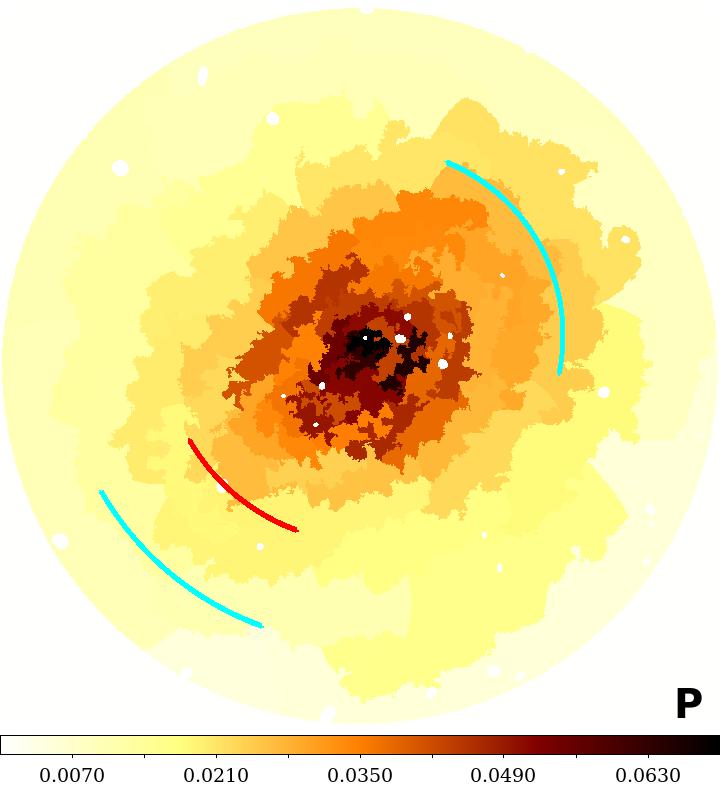}
\caption{{\bf From top left, clockwise:} The temperature (keV), pseudo density (cm$^{-3}$ $l^{-1/2}$), pseudo pressure (keV cm$^{-3}$ $l^{-1/2}$) and pseudo entropy (keV cm$^{2}$ $l^{1/3}$) maps of Abell 2219, out to 200 arcseconds ($\sim$720 kpc). These figures display 3,600 counts per bin, which at these temperatures allows us to place $<$ 15 per cent uncertainties in the temperature. Relative abundances are fixed to 0.3 times the solar value. The maps are the same scale, with N up and E to the left. The positions of the BCGs are marked with blue crosses on the temperature map and the positions of the surface brightness edges discussed in Section \ref{imaging} are marked with cyan arcs. A further density discontinuity (see Fig. \ref{radial2}) is marked in red. Montecarlo maps showing the uncertainty and covarience in the temperature and density maps are located here: \protect \href{http://www.web.stanford.edu/~rcanning/a2219.html}{http://www.web.stanford.edu/$\sim$rcanning/a2219.html}.  \label{thermo_large}}
 \end{center}
\end{figure*}

The mosaiced, exposure-corrected X-ray image, with a 2D elliptical beta model fitted and subtracted (double beta models are not formally better fits), is shown in the top panel of Fig. \ref{unsharpmask}. This reveals two surface brightness edges (marked in cyan on the image) along the projected merger axis on either side of the centroid. The centroid of the 2D beta model is SE of the brightest region of X-ray emission between the two dominant galaxies and is marked with an `X'.

The bottom panel of Fig. \ref{unsharpmask} shows an unsharp-masked X-ray image (made by subtracting an image smoothed by a Gaussian with $\sigma=35$ from an image smoothed by a Gaussian with $\sigma=9$) of the core, overlaid on a three colour optical Subaru Suprime-Cam image. A complicated structure is revealed in the core with two bright regions of X-ray emission associated with both the brightest cluster galaxy (BCG 1) and the second largest galaxy (BCG 2) of the A2219 system. These galaxies have a projected separation of approximately 50 arcseconds ($\sim$180~kpc). \cite{boschin2004} find a line-of-sight velocity difference between BCG 1 and 2 of 1,320 \kmps.

\section{Thermodynamic maps}
\label{thermo}

In Fig. \ref{thermo_large} we show projected maps of temperature (in units of keV), pseudo density (cm$^{-3}$ $l^{-1/2}$), pseudo pressure (keV cm$^{-3}$ $l^{-1/2}$) and pseudo entropy (keV cm$^{2}$ $l^{1/3}$) for Abell 2219 (where $l$ is normalised to 1 Mpc). The {\sc contbin} routine \citep{sanders2006c} was used to bin the 0.6--7.0~keV images, with point sources masked, to a signal-to-noise of 60, requiring $>3,600$ background-subtracted counts per bin. The spectra were fitted between 0.6--7.0 keV\footnote{Higher energy cut-offs were also used but did not change the results.} with an {\sc apec} thermal model \citep{smith2001}, with a Galactic equivalent column density of N$_{\mathrm{H}}=1.76\times$10$^{20}$~\pcmsq \citep{kalberla2005}. At the high systemic temperature of Abell 2219 ($\sim10$~keV) this binning allows us to place $\lesssim15$ per cent uncertainties on the projected temperature measurements and $\lesssim5$ per cent uncertainties on the pseudo density measurements in each bin. 

The metallicity was fitted using larger regions with 20,000 counts per bin. However, due to the high cluster temperatures, 20,000 counts per bin only allows for 30 per cent statistical errors on the abundance. 
Within the uncertainties, the fitted abundance values are consistent with a constant value of 0.3 across the system, with only a slight ($\sim2\sigma$) indication for enhanced metallicity at the position of BCG1 (and the X-ray `peak'). The maps shown in Fig. \ref{thermo_large} have relative abundance fixed to 0.3 times the solar value. 

The top left hand plot of Fig. \ref{thermo_large} reveals a `horseshoe' of high temperature gas, which surrounds the core, and is oriented towards the NW. This feature, with a temperature $>$14.5~keV, extends over 8 independent regions in the map. In the NW, the projected location of the high temperature gas is roughly coincident with the NW surface brightness edge discussed in Section \ref{imaging} (marked in cyan on Fig. \ref{thermo_large}) and coincides with a the shock front first reported by \cite{million2009}. The entropy and density maps both have relatively sharp features at this location, with the entropy rising just beyond the edge. SE of the core, a tail of low temperature gas is apparent. The tail is coincident with low entropy, low pressure gas. 

The entropy map is asymmetric: it has a highly elongated structure, orientated NW-SE, that rises more quickly in the NW. The pressure map is `rounder' although the pressure gradient drops more rapidly towards the SE. Regions of high pressure are observed perpendicular to the projected merger axis. 

\begin{figure}
 \begin{center}
  \includegraphics[width=0.45\textwidth]{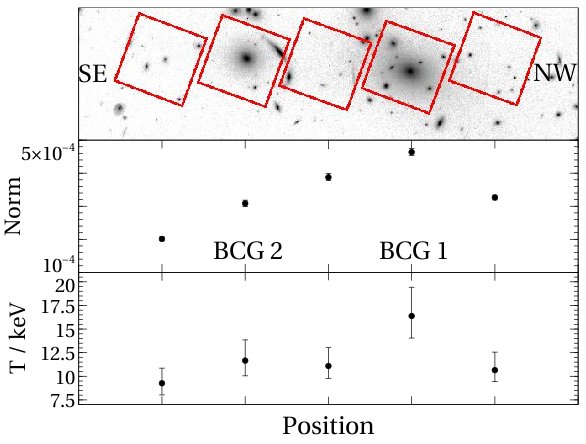}
\caption{The {\sc xspec} normalisation and temperature in 5 equal sized regions spanning the core. The regions are shown in the top panel. The second and fourth regions coincide with the second and first BCGs, respectively. The density increases rapidly to a peak on the larger, dominant BCG from the NW, then declines slowly to the SE due to the second `peak' from the smaller cluster. The gas temperature is high at the dominant BCG (BCG1), while at the second BCG the temperature is consistent with the neighbouring regions within the statistical uncertainties. \label{core_boxes}}
 \end{center}
\end{figure}

The core is complex, with a higher density and lower entropy than the surrounding gas but also with a high temperature and pressure.  The positions of the two dominant galaxies, are indicated on the temperature map with blue crosses. They both coincide with high temperature regions. To establish whether these high temperature peaks are significant, we extract point source removed spectra from 5 regions, of equal size, spanning the core and orientated parallel to the projected merger axis (see Fig. \ref{core_boxes}). We cannot provide strong constraints on the metallicities or any multiple temperature structure of the gas. However, single temperature {\sc apec} fits show that the surface brightness peak, coincident with BCG 1, is significantly hotter and denser than its surroundings. The density declines more slowly towards the SE of BCG 1 due to the contribution associated with BCG 2 and the second subcluster. At BCG 2, the temperature is high but within the statistical uncertainties of the neighbouring regions.

\section{Profiles}
\label{profiles}

The imaging analysis and thermodynamic maps show two clear surface brightness edges at large radii ($>400$~kpc) along the direction of the projected merger axis. There are also complex, high pressure, high temperature regions surrounding the core. We examine these inner and outer features in more detail in this section.

Whilst the intermediate, 1.8--3.0~keV, band provides good density sensitivity, independent of temperature, for high temperature gas, the 0.6--7.0~keV full-band image is also relatively insensitive to variations in the gas temperature. Due to the improvements in signal gained by using the full band we therefore choose to make our surface brightness profiles using the full band. The counts are extracted separately on the merged counts image and merged background image. These are divided by their respective exposure maps and the background counts are subtracted. Uncertainties are propagated in quadrature.  

Aside from where specifically stated, the thermodynamic profiles presented are all deprojected profiles that assume spherical symmetry. We do not combine the spectra across different chips; instead we fit the data from each OBS ID and chip simultaneously in {\sc xspec}.

\begin{figure*}
 \begin{center}
  \includegraphics[width=\textwidth]{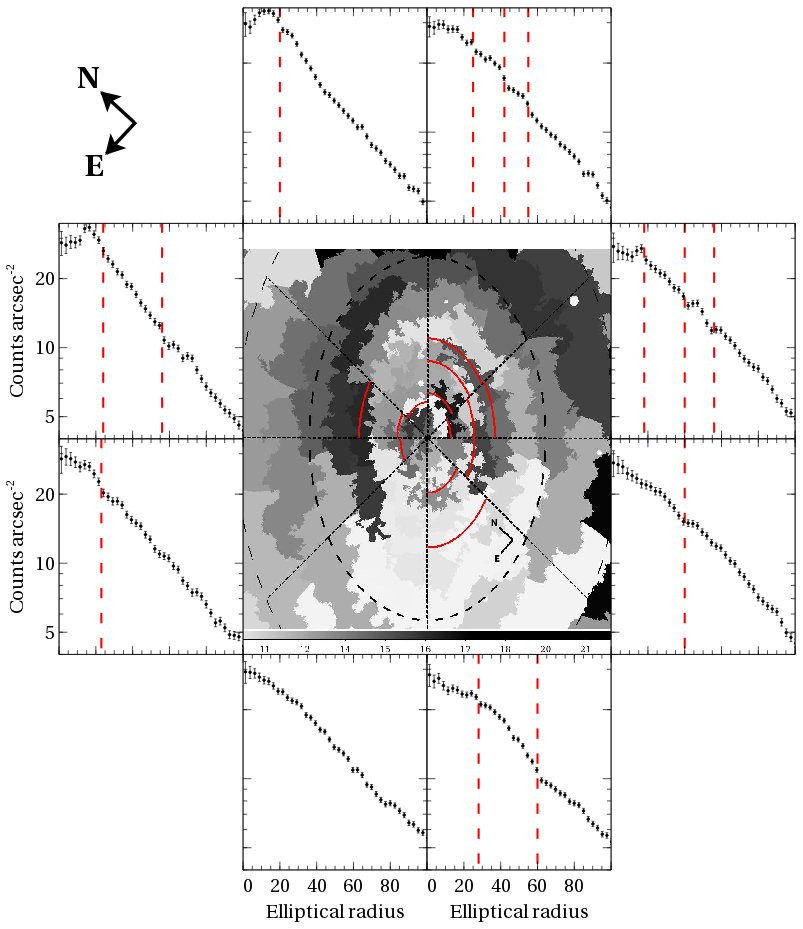}
\caption{Surface brightness profiles extending to 100 arcseconds ($\sim$350 kpc) elliptical radius from the X-ray centroid of Abell 2219. The figure is orientated such that projected merger axis (NW to SE) is vertical with NW being up. The core surface brightness profiles are complex with many edges observed. The projected temperature map is shown in the centre and many surface brightness edges correlate with regions of projected high temperature. To guide the eye, we have marked the obvious surface brightness edges with red dashed lines on the profiles and the corresponding arc on the temperature map. The centre, major and minor axis ratio and position angle are from the best fit elliptical beta model. This is centred at RA 16$^{h}$40'20.173' and Dec $+$46$^{d}$42'30.60'' with position angle $\theta=40.8^{\circ}$ W from N. \label{core_prof_1} \label{core_prof_2}}
 \end{center}
\end{figure*}

\begin{figure}
 \begin{center}
  \hspace{0.95cm} \includegraphics[width=0.35\textwidth]{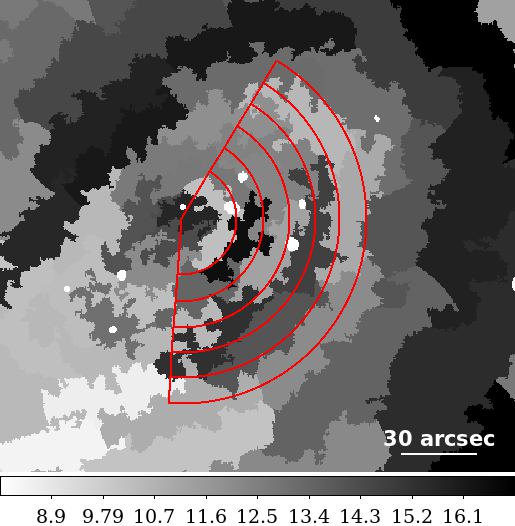}
  \includegraphics[width=0.43\textwidth]{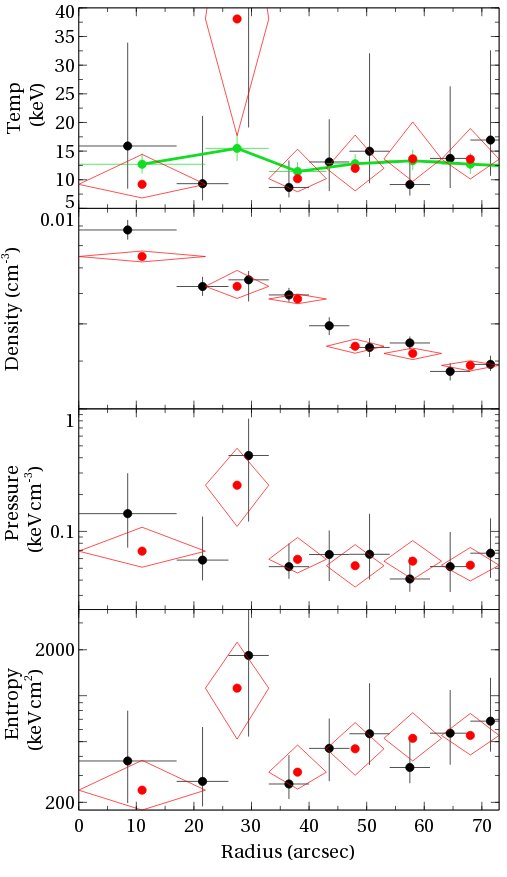}
\caption{{\bf Top:} Temperature map of the core of Abell 2219. There are a minimum of 3,600 counts per bin. {\bf Bottom:} Deprojected temperature, density, pressure and entropy profiles in the core of Abell 2219. The red points are extracted from the annular regions shown in the top panel (red), which have a minimum of 8,000 counts per region. The black points are extracted from the same sectors but with 5,000 counts per region. The green points show the projected temperatures in regions with 8000 counts. All abundances are fixed at 0.3 solar. There is weak evidence for temperature and density discontinuities in an arc roughly $\sim$20--30 arcseconds from the core, which coincides with the regions of high projected temperature in the map above. Three surface brightness edges exist at $\sim$20, 40 and 55 arcseconds.  \label{inner}}
\end{center}
\end{figure}

\subsection{The core}

Fig. \ref{core_prof_1} shows the surface brightness profiles of 8 regions surrounding the core, extracted in 2 arcsecond radial bins. The centre and position angle of the elliptical profiles are those of the fitted centroid of the best fit elliptical 2D beta model discussed in Section \ref{imaging}. 

The steepest surface brightness profiles are in the N and NE, offset slightly from the direction of the projected merger axis. SE of the core, the surface brightness exhibits a distinctly shallower profile. A series of edges surround the core and correspond well with regions of high projected pseudo-pressure and high temperature. These edges are shown on the profiles in Fig. \ref{core_prof_1}, indicated by dashed red lines, and their positions in 2D are indicated on the projected temperature map, in the center of the figure.

The surface brightness edges in the core are sharp, narrow features. It is therefore difficult to examine the spectral properties of these regions. The requirement of $\sim$5000 counts to measure the deprojected temperature limit us to using relatively large regions from which to extract spectra. Fig. \ref{inner} shows the spectral properties extracted in regions towards the W of the core where the highest projected `inner' temperatures and pressures are observed. At least three consecutive surface brightness edges are observed in the same direction at radii of $\sim$25, $\sim$40, and $\sim$55 arcseconds (see Fig. \ref{core_prof_1}). An increase in both projected and deprojected temperature is observed at $\sim$30 arcseconds ($\sim$108 kpc) from the core. Discontinuities or edges in the surface brightness, density, pressure and entropy are also observed, indicative of a shock.

\subsection{The outer edges}
\label{outere}

\begin{figure*}
 \begin{center}
  \includegraphics[width=0.4\textwidth]{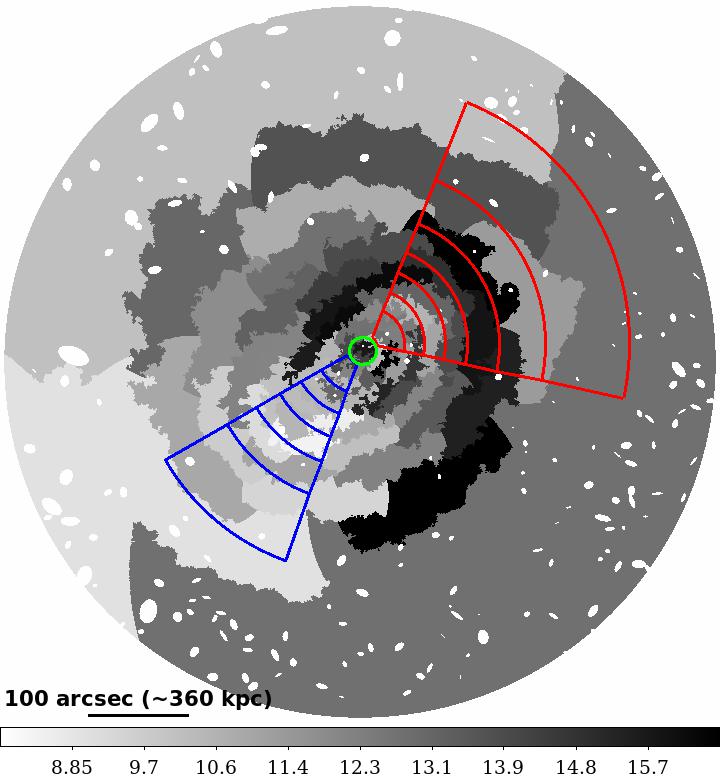}
  \includegraphics[width=0.8\textwidth]{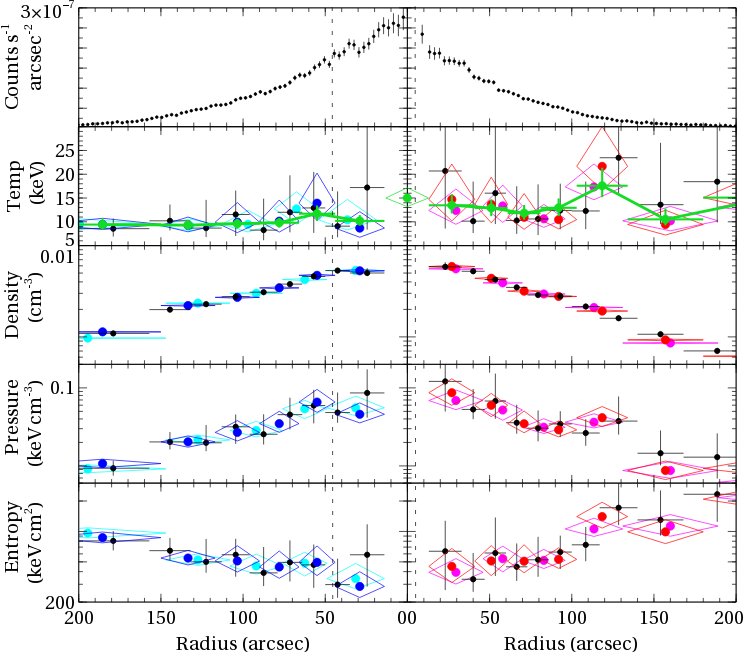}
\caption{The surface brightness, deprojected temperature, density, pressure and entropy profiles out to 200 arcseconds. Regions with $\gtrsim$8000 counts are indicated in blue and red on the top image and in the bottom profiles. The black points have $\gtrsim$5,000 counts per region, and the magenta and cyan diamonds have $\gtrsim$10,000 counts per region. The green central temperature and the green points in the temperature profiles are projected temperatures from the 8,000 count annular regions. The abundances are fixed at 0.3 solar. A series of edges discussed in Fig. \ref{core_prof_2} can be seen within 50 arcseconds in the surface brightness profiles and a clear temperature jump indicating a shock front, likely the bow shock, is observed at $\gtrsim$100 arcseconds in the NW. The SE is cooler and the profiles are smoother. The position of BCG 1 and BCG 2 are indicated with dashed lines in the lower panels. \label{radial1}}
 \end{center}
\end{figure*}

Initially, we select by eye sectors which trace the edges observed in Fig. \ref{unsharpmask} from which to extract surface brightness profiles and map the spectral properties. The surface brightness and thermodynamic profiles in these sectors are shown in Fig. \ref{radial1} in regions of 5,000, 8,000 and 10,000 counts per bin.

The thermodynamic profiles in the NW show a significant temperature jump 100-140 arcseconds from the core. The temperature discontinuity is accompanied by a density, entropy and pressure discontinuity. The direction of the temperature and density jumps indicate a shock front. 

In the SE, a possible temperature increase is observed at $\sim$50 arcseconds in close proximity to a surface brightness edge and the location of BCG 2 (dashed line). A third possible density discontinuity is also apparent at a radius of $\sim$150 arcseconds.

\begin{figure*}
  \begin{center}
   \includegraphics[width=0.42\textwidth]{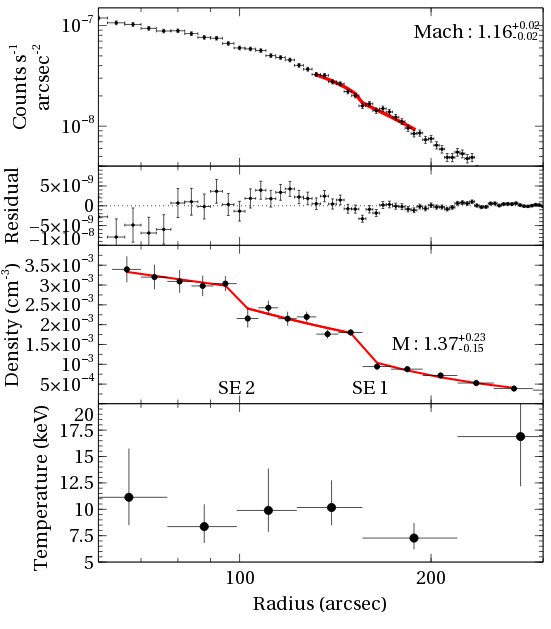} 
   \includegraphics[width=0.42\textwidth]{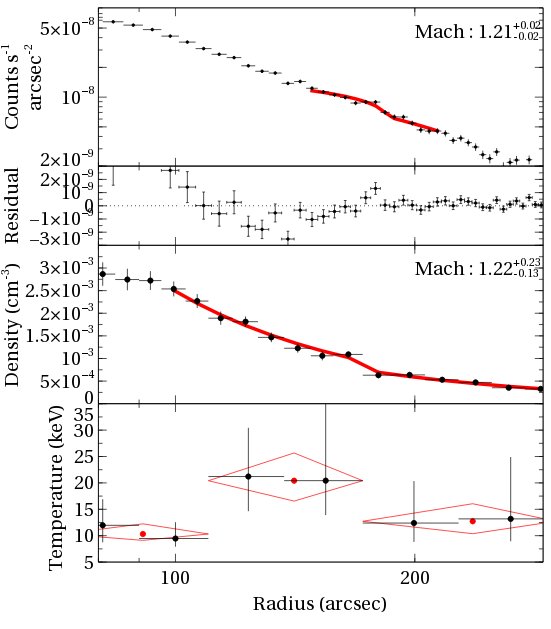} 
 \caption{Surface brightness and thermodynamic profile through sectors optimised to enharnce the surface brightness edges (error bars are 90\%). The {\bf left} hand panel shows the profiles across the SE edge and the {\bf right} hand panel across the NW edge. {\bf Top:} The X-ray surface brightness profiles and a residual profile after subtraction of a single $\beta$-model fit. The residuals are shown here as deviations from the smooth $\beta$-model, indicating the location of the edges is more easily observed in the residual profile. {\bf Middle:} Deprojected density profile (2,000 counts per region). Two discontinuities are observed in the SE. While not clear from the surface brightness profile itself the discontinuity does correspond to a deviation from the smooth profile observed in the residual profile. {\bf Bottom:} Deprojected temperature profile (black - 6,000 counts per region; red - 12,000 counts per region).  Fits to the density discontinuities and surface brightness edges are shown with solid red lines. \label{radial2}}
  \end{center}
 \end{figure*}

We further refine our surface brightness extraction regions by dividing the NW and SE sectors into three, measuring the strength of the surface brightness `edge' in each of the three sectors and varying the position angles over which they are extracted. Though we have tried to mitigate them, uncertainties in estimating the ellipticity of the edge and the centring of the extraction regions will exist. These effects blue the edges leading to less contrast in surface brightness and thermodynamic features (unfortunately our signal-to-noise is not high enough at the edge to bring an edge model directly to our X-ray image and fit for these parameters). Our `sharpest' surface brightness profiles are shown in Fig. \ref{radial2} (the width of the annuli are chosen to be a minimum of 4 arcsec or alternatively a minimum of 100 counts per region).

We extract spectra from the same sectors with the requirement that each of the annuli from which spectra were extracted must have a minimum of 2,000 counts per region in the 0.6--7.0 keV band (calculated from the background subtracted, point source removed images). Since, the temperature is not well constrained with only 2,000 counts, we tie the temperature in adjacent regions to a minimum of 6,000 (black points) counts per region. We use the surface brightness profiles to ensure the temperatures are not tied across a surface brightness edge\footnote{Deprojecting the profiles in low surface brightness regions can introduce oscillatory noise into the neighbouring bins. To mitigate this we used the hard band (4.0-7.0 keV) image to determine the radii to which we should extend our measurements. A 12 keV APEC model at the redshift of our cluster has 1.4 per cent of the counts in this energy range compared with the 0.6-7.0 keV band. We therefore require our profiles to have $>$50 counts in the hard band}. The profiles are shown in Fig. \ref{radial2}. 

The shock front is clearly seen towards the NW ($\sim150$'') and, surprisingly, two density discontinuities are seen in the SE. The farthest ($\sim170$'') corresponds to the edge observed in the unsharpmasked image (Fig. \ref{unsharpmask}), but an additional discontinuity, not seen in the surface brightness profile, is also observed at $\sim100$ arcseconds.

To estimate the edge properties, we model the density on either side of the discontinuities as a power-law of the form:
\begin{equation}
 n_{i}=N_{i}(r/R_{f})^{-\alpha_{i}},
\end{equation}
where $n_{i}$ is the density, $\alpha_{i}$ is the power law index, $N_{i}$ is the normalisation at each side, $i$, of the discontinuity and $R_{f}$ is the position of the discontinuity. This can be fit directly to the deprojected density profile or projected and fit to the surface brightness profile (e.g. \citealt{owers2009}). To model the surface brightness we make the assumptions of spherical symmetry, that the underlying beta model can be neglected and that the surface brightness is not sensitive to the temperature variations about the shock (as mentioned in Section \ref{profiles}, at the high temperatures of Abell 2219 and in the X-ray energy-band considered, this is a reasonable assumption).

The best fits to the surface brightness profiles and to the deprojected density profiles are shown by the red lines in Fig. \ref{radial2}. The uncertainties on the fits are found using an monte-carlo approach and the errors quoted on Fig. \ref{radial2} are 95\%. However, it should be noted that changing the region over which the fit is performed influences the best fit. In practice, we must use regions which are large enough to have many data points to fit to either side of the jump but small enough that the assumption of power-laws in density is valid and that other sharp features in the surface brightness profiles are not included. The regions used for the fits can be seen on Fig. \ref{radial2}. To estimate the systematic uncertainty we perform the fit over wider and narrower regions (3 data points either way); the systematic uncertainties on the Mach numbers are typically 0.15.
The results from these fits indicate:
\begin{enumerate}
\item {\bf SE 1 shock front ($\sim$155'', 556~kpc):} A fit to the surface brightness edge gives $M=1.16$ whilst the density discontinuity gives $M=1.37$. Taking into account the systematic uncertainties the Mach numbers are consistent.  A simple F-test rules out a straight line fit to the density discontinuity with 99\% probability, and $\Delta T=\frac{T_{2}}{T_{1}}$ at the interface is also inconsistent with unity of temperature to greater than 0.01\%. The best fitting value of $\Delta T=0.7$, indicates a Mach number of $M=1.2$, and the sign indicates a shock front. A Mach number of 1.37 from the density discontinuity leads to a velocity of the shock of $\sim$2,350~\kmps\ for a $T\sim10$~keV gas. The lower Mach number of 1.16 reduces the velocity to $\sim$2,000~kmps.
\item {\bf SE 2 cold front ($\sim$100'', 359~kpc):} 
The projected density model gives a reasonable fit to the surface brightness, but, a simple F-test cannot reject the null hypothesis that no discontinuity in density exists or that no edge exists in the surface brightness profile. However, the temperature measurements across the discontinuity are inconsistent at the 3\% level, with the best fit $\Delta T=1.15$. 
\item {\bf NW shock front: ($\sim$166'', 595~kpc)} The density discontinuity is inconsistent with no jump at the 99\% level. A fit to the density discontinuity gives $M=1.22$, consistent with the surface brightness edge. This leads to a velocity of the shock for a $T\sim10$~keV gas of $\sim$2100~\kmps. The projected distance from the centroid is 166 arcsec (595~kpc), which allows us to estimate the time since core passage of the shock as 0.26~Gyrs. The best fit value of $\Delta T = 0.63$, suggests a higher Mach number of 1.4, the temperatures are inconsistent with no variation at the 0.01\% level. However, we cannot rule out a smaller temperature jump consistent with $M=1.2$.
\end{enumerate}

\subsection{Radio-X-ray connection}
\label{radioxrayconnection}

\begin{figure}
 \begin{center}
  \includegraphics[width=0.39\textwidth]{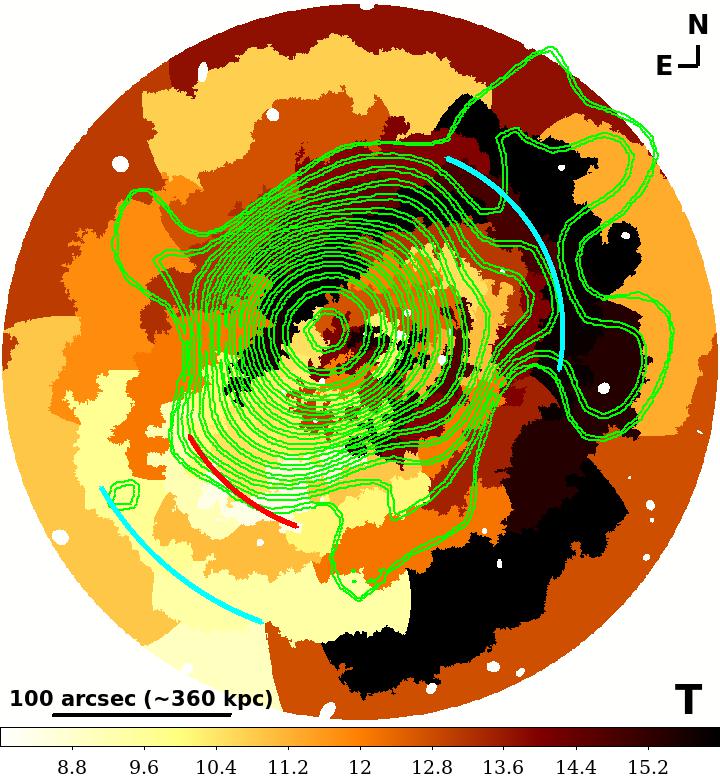}
  \includegraphics[width=0.39\textwidth]{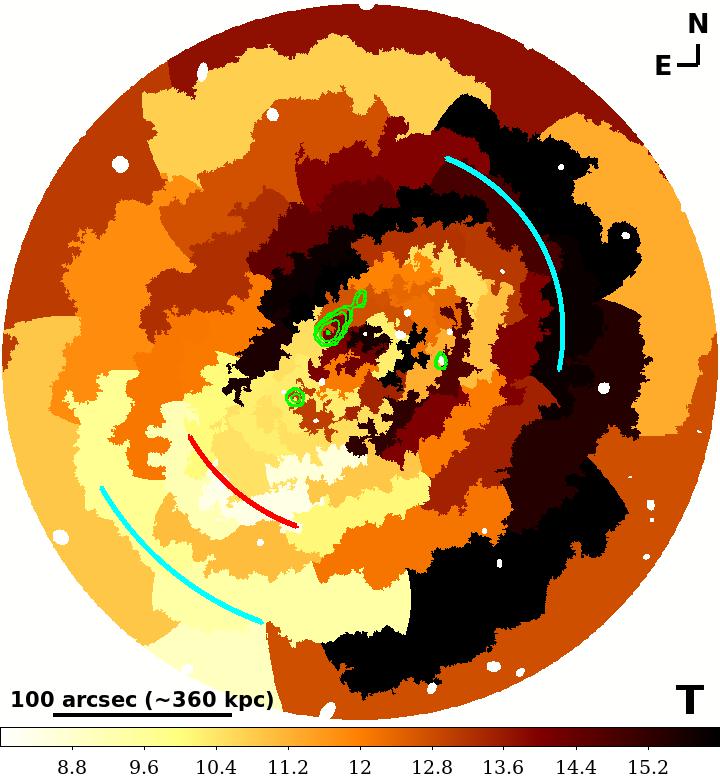}
  \includegraphics[width=0.39\textwidth]{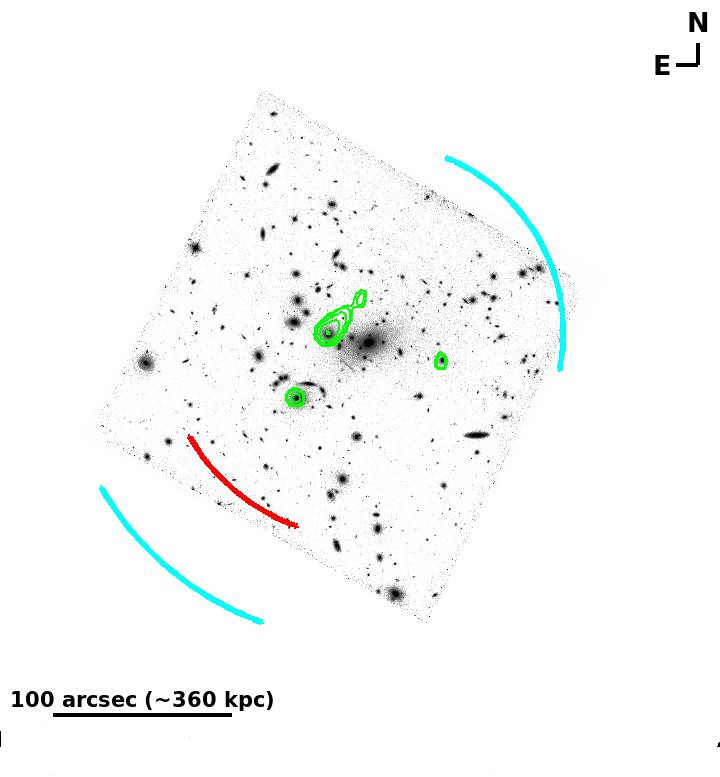}
\vspace{-0.7cm}
\caption{X-ray projected temperature map overlaid with contours of diffuse radio emission (325~MHz radio brightness contours reproduced from \protect \citealt{orru2007}) {\bf (top)} and VLA 1.4 GHz emission indicating the radio galaxies {\bf (middle)}. {\bf Bottom:} The HST optical imaging on the same scale. The position of the `outer' edges, discussed in Section \protect \ref{outere}, are indicated by a red arc (cold front) and cyan arcs (shock fronts).  \label{xray_radio}}
 \end{center}
\end{figure}

Fig. \ref{xray_radio} shows the morphology of the diffuse (top panel, 325 MHz VLA radio contours reproduced from \citealt{orru2007}) and point source (middle panel, 1.4 GHz VLA radio contours) radio emission overlaid on the X-ray projected temperature map. The location of the fronts are labelled.
Interestingly, the radio halo appears to be bounded by the shock front in the NW and the cold front in the SE (Fig. \ref{xray_radio}). Merger radio halos are thought to be generated by turbulence (e.g. \citealt{brunetti2001, petrosian2001}) while merger radio relics, or radio `gischt' \citep{kempner2004}, are located at the periphery of clusters and associated with particle acceleration at shock fronts (e.g. \citealt{ensslin1998}). Despite these definitions, there are examples of radio halo edges coincident with X-ray shocks (e.g. Coma cluster, \citealt{brown2011}; Abell 520, \citealt{markevitch2005}; Abell 665, \citealt{govoni2004, feretti2004}; Bullet Cluster, \citealt{markevitch2002, shimwell2014}).

\subsubsection{NW and SE X-ray shock fronts}

The morphology of a radio surface brightness edge corresponds well to the high temperature X-ray gas at the NW shock front. However, while the high temperature gas has a clear horseshoe shape the radio emission has a more limited extent.
These differing extents may be a reflection of the increased gas density along the projected merger axes (Fig. \ref{thermo_large}), the distribution of the pre-existing electron populations or an indication of where turbulence is most efficiently generated behind the passing shock. 

The surface brightness and density discontinuities of the NW shock are consistent with a low Mach number of $\sim$1.2 while the temperature favours $M\sim$1.4. If the edges suffer from projection effects the density Mach numbers could be underestimated (e.g. \citealt{markevitch2001, ogrean2013}).
The SE reverse shock also exhibits a low Mach number of $1.1-1.4$, with the temperature jump in this case providing a lower Mach number than the density discontinuity. However, the temperature measurement is likely contaminated by the colder tail of stripped gas emanating from the disrupted sub-cluster core. Unlike the NW shock front the SE shock front is not associated with strong diffuse radio emission.

The morphological connection between the radio halo and NW shock front in A2219 is strengthened by the observation that the spectral index steepens, from $\sim$0.8, in the halo, to $\sim1.4$, at the location of the front (see \citealt{orru2007}, their Fig. 9). If we make the assumption that the radio emission near the shock is accelerated by Fermi acceleration of the shock front the expected radio spectral index is $\alpha=((2(M^{2}+1)/(M^{2}-1))-1)/2  + 1/2$, where $M$ is the shock Mach number \citep{blandford1987}. For a spectral index of 1.4, we require Mach numbers $>2.5$, significantly higher than measured. The density of the shocked gas would need to be 2.7 times the density in the post shock region whereas we measure 1.5 (a density ratio of this amount could occur if the post shock gas takes less than 20\% of the volume of our measured bin). This argues against direct acceleration, as does arguments about the radiative lifetime of the electrons ($\sim10^{8}$~yrs) which is several times shorter than the shock crossing time in this system (e.g. \citealt{brunetti2003}). Moreover, we see that the radio emission extends beyond the shock front. 
Re-acceleration of a existing population of energetic electrons may be occurring: these electrons could originate from an existing radio halo or have been generated by the three known radio galaxies 
\citep{owen1992}. The lack of radio emission associated with the SE shock may indicate that turbulence is low behind the in-falling subcluster and not yet widespread throughout the merging system.

The spectral index map of the radio emission appears not to be smooth, with regions N and W of the core steeper than the S (localised variations in the spectral index of the Bullet Cluster were also observed by \citealt{shimwell2014}). This may indicate that the halo is still forming and the diffuse radio emission is being varied locally by the proximity to shocks and the amount of turbulence. N and W of the core of A2219 we detect multiple edges with the characteristics of shocks (see Section \ref{hot_core}). Although these have very low measured Mach numbers they may influencing the radio emission in these central regions.

\subsubsection{SE X-ray cold front}

In the SE, the radio edge coincides very well with the SE cold front providing further evidence that this is a real feature. The radio emission sits inside the edge, as is observed in `sloshing' cold fronts in relatively relaxed systems (e.g. \citealt{mazzotta2008}). The abrupt edge to the halo supports the hypothesis that this `remnant tail' gas is not yet mixed with the ICM. The spectral index map of \cite{orru2007} suggests that the spectral index steepens towards the SE boundary ($\sim1\pm0.15$) where we find the X-ray cold front. We note here the uncertainties on this steepening are large. 

In merger cold fronts the radio emission is amplified outside of the front, where the magnetic field is being swept up and stretched out, by the motion of the front through the hotter and less dense medium \citep{lyutikov2006}.
In cold fronts generated by sloshing of dense X-ray cores the shearing motions both amplify the magnetic fields tangential to the front and generate turbulence {\it inside} the front. This turbulence can re-accelerate the electrons and will result in a steeper radio index \citep{zuhone2013}. It's not clear whether the sub-cluster hosted a radio-halo pre-merger. However, BCG 2, which is associated with the sub-cluster, is a radio galaxy (bottom panel Fig. \ref{xray_radio}), and may have provided the population of electrons to be re-accelerated by the sloshing of its core as it passes through the ICM of the main cluster.

The survivability of a dense core under these conditions is not obvious. While the sloshing core and magnetic draping, at both the leading edge and due to the sloshing, may help confine the dense core gas and prevent much stripping and mixing, turbulence is also generated within the front which would act to disrupt the field and promote Kelvin-Helmholtz instabilities.

\subsection{inverse compton}

Inverse Compton (IC) emission from up-scattered CMB photons should be generated by the same population of relativistic electrons producing the radio halo. This non-thermal X-ray emission can in principle be detected as a power-law component or soft-excess at X-ray wavelengths (e.g. \citealt{million2009}). The radio halo emission peaks in the core and the inferred high core X-ray temperatures could be indicative of an IC component to the spectrum. Following \cite{million2009}, we test this possibility by both allowing N$_{\mathrm{H}}$ to vary spatially, and by fitting an additional power-law component in regions with at least 15,000 counts. We find only one region, near the peak of X-ray emission, where N$_{\mathrm{H}}$ is significantly different from the Galactic value. If N$_{\mathrm{H}}$ is fixed to the Galactic value and an additional power-law component is fit, the best fit power law index requires an un-physical spectral index of $\alpha=-0.1$ (the radio halo synchrotron index measured by \citealt{orru2007} is $\alpha=0.8-1.0$). Whilst a two-temperature gas, with fixed abundance, does not offer a significant improvement in the goodness of fit to our data it is likely that the variation in N$_{\mathrm{H}}$ is caused by a multiphase gas associated with the dense core of the dominant subcluster.

\section{Discussion}
\label{discussion}

The Abell 2219 system is undergoing a merger of two galaxy clusters in the NW-SE direction.
The morphology of the X-ray emission is elongated in this direction and three density discontinuities are observed along the projected merger axis. 
The system has a complicated core structure with multiple surface brightness edges and a long tail of stripped cooler gas to the SE.

The most striking feature of the merger is a horseshoe of high temperature gas at 166 arcsec (595 kpc) towards the NW (Fig. \ref{thermo_large}), the leading edge of which coincides with a shock front (Fig. \ref{radial1} and \ref{radial2}). 
In the reverse direction, at 155 arcsec (556 kpc), from the centroid, we detect, with $>$99\% confidence in a temperature jump, a second shock front. The NW shock is associated with diffuse radio emission, having a steeper spectral index than the remainder of the halo, yet the SE shock shows no evidence for associated radio emission.

A probable cold front is observed in the SE wake of the stripped sub-cluster core. However, neither the surface brightness profile nor the deprojected density profile can reject at high confidence the null hypothesis that no jump exists (although the coincidence of the feature with a radio edge adds evidence for the discontinuity being real, see Section \ref{radioxrayconnection} and Fig. \ref{xray_radio}). 

Recently, \cite{roediger2014}, simulated the disruption of the atmospheres of elliptical galaxies falling into clusters. The authors found much of the gas, forming the wake, remains unmixed as a `remnant tail' for long periods and, within this wake, sloshing cold fronts are expected due to the `push' of the ICM. Behind the `remnant tail' is a deadwater region and stripping and mixing of gas only occurs beyond this deadwater region. Whilst these simulations were for individual galaxies, we may be observing a similar process in the stripped atmosphere of the sub-cluster.

The long, low entropy and temperature wake in the SE suggests there has been little mixing of the tail gas with the surrounding hotter ICM. The pressure map is smooth beyond the tail edges indicating pressure equilibrium is maintained with the surrounding gas. The feature observed at 359 kpc might be a sloshing cold front, within the remnant tail, or could indicate the end of the remnant tail, where the deadwater region occurs. 

\subsection{Hot core and multiple surface brightness edges}
\label{hot_core}

\cite{million2010} identified a hot core in the A2219 system. Our observations confirm a temperature increase in several regions in the core. Notably, both BCGs, appear to be associated with high temperature regions, with the dominant BCG 1 associated with the highest temperatures ($\sim$15~keV), and densest gas (0.01\pcmcu), indicating that the merger shocks are able to penetrate the central regions.

A series of `inner'  surface brightness edges are also observed close to the centroid of the system (within 50'' or $\sim$180 kpc). Some edges are likely cold fronts from the sloshing cores (e.g. `bridges' and `edges' observed in the simulations of \citealt{poole2006}) but a few of these edges are similarly associated with high temperature, high pressure regions and are likely driven by shocks around the core. We are unable to precisely constrain the deprojected temperatures and densities in these regions due to the combination of the small widths of the shock regions and the high temperature gas. However, from the surface brightness discontinuities, we estimate Mach numbers of $\sim$1.15. 

BCG 1 is not a strong radio source (see Fig. \ref{xray_radio}, bottom panel), and there are no indications in our X-ray surface brightness of `holes' resulting from bubbles blown by AGN jets, so the shocks are more likely a result of the dynamics of the merger rather than AGN driven shocks. The projected pressure maps show the dominant disruption occurring at the sides of the core, perpendicular to the projected merger axis. This is where one might expect the shear flow is largest and the gas less protected and more easily stripped \citep{roediger2014}. However, from optical spectroscopy, \cite{boschin2004} propose that the main A2219 core has also suffered the repeated in-fall of many small groups. In this scenario the surface brightness edges may be a consequence of many shocks experienced by the dense core due to the in-fall.

\subsection{Anatomy of a merger}

The X-ray morphology, the tail of cool gas and the outer surface brightness edges suggest the system is undergoing a significant merger event, with the main component along the NW-SE axis. This is consistent with  
the direction of elongation of the galaxy population. However, \cite{boschin2004} also present evidence for a significant line-of-sight velocity gradient ($\sim$1000~\kmps) in the E-W direction; from this they suggest the merger must have a component perpendicular to the plane of the sky, and that a $\sim$45$^{\circ}$ angle is the most likely. The two BCGs also have a significant line-of-sight velocity difference ($\sim1300$\kmps).
If the merger is not exactly in the plane of the sky, projection effects would result in us under-estimating the shock Mach numbers. A higher Mach number would alleviate tensions between the radio edge emission and low Mach number NW shock.

Simulations show that a forward propagating shock can have a `horseshoe' structure around core passage, with these features being wiped out after $\sim$1 Gyr \citep{ricker2001, poole2006, skillman2008}. The strongest shock is observed leading the less massive sub-cluster.
Our projected temperature map shows a similar configuration, supporting the hypothesis that this is a very recent merger ($\sim$at core passage) - by non-equal mass - clusters, with the smaller cluster approaching from the SE (strong lensing also suggests the smaller sub-cluster is in the SE; \citealt{smith2005}). As it seems unlikely that the smaller subcluster would harbour the larger BCG the merger is probably being observed just before closest approach. At a projected distance of 595~kpc, the shock velocity translates to a time since core passage of the shock of $\sim0.26$~Gyrs, supporting the scenario of a youthful merger.

\section{Conclusions}
\label{summary}

Abell 2219 is a system in the throes of a violent merger, with the dominant component of the motion in the NW-SE direction. Both shock fronts and cold fronts in this system appear to be influencing the radio halo morphology and structure. The key results from our {\it Chandra} analysis are:
\begin{enumerate}
 \item The projected temperature maps show that the cores are surrounded by a high temperature `horseshoe'. In the NW, this is coincident with the surface brightness edge and density discontinuity which define the outer shock front. The shock front is coincident with a radio halo edge with a steeper spectrum than the remainder of the halo. The shock Mach number is $M\sim1.2$, leading to a velocity of $\sim$2100\kmps and time since core passage of the front of 0.26 Gyr.
 \item A reverse shock with similar low Mach number is observed in the SE. This shock is not associated with any radio emission indicating that shock induced turbulence is not yet widespread in the cluster. 
 \item Simulations of cluster mergers at core passage show similar `horseshoe' temperature structures and indicate that this is a early stage merger between non-equal mass ratio clusters. The youthful nature of the merger is supported by the positions of the BCGs; the first and second BCG are $\sim$180~kpc apart and have probably yet to cross paths.
 \item We present the first evidence for a sloshing cold front observed in the `remnant tail' of an in-falling sub-cluster. The cold front is observed in the SE and bounds the diffuse radio halo. The spectral index may steepen at this boundary indicating the cold front is amplifying the field at this location.
 \item Both the X-ray cores associated with the BCG 1 and 2 are `hot' and multiple surface brightness edges surround them. These edges are likely of both shock and cold front origin and correspond to Mach numbers of $\sim$1.15.
  
\end{enumerate}

The shock fronts in A2219 are not at great distances (the outermost shock being $<$600 kpc from the centroid and probable inner shocks within 180 kpc), so they have yet to propagate to the very low surface brightness regions of the system. As such A2219 presents an interesting opportunity to study the detailed connection of the thermal X-ray population and the non-thermal radio synchrotron emission. A population of relativistic electrons must exist in the system, generated by the three known radio galaxies, and the merging ICM is likely to be highly turbulent. However, the correspondence of the radio emission with the NW X-ray shock, and the spectral index variations in the core, where we see evidence for multiple shocks is tantalizing; low Mach number shocks may be more efficient at promoting synchrotron emission than currently thought. The tail of low temperature X-ray gas, sharp radio edge at the location of the cold front and lack of radio emission associated with the reverse shock suggest there is little mixing and turbulence behind the in-falling core. High resolution X-ray data combined with radio imaging and polarisation data should enable us to determine the role the shocks and cold fronts are playing in compressing the cluster fields and re/accelerating the electrons and ions.

\section{Acknowledgements}
REAC. wishes to thank Norbert Werner and Irina Zhuravleva for interesting discussions.
Support for this work was provided by NASA through grant number GO2-13143X.
SWA. acknowledges support from the U.S. Department of Energy under contract number DE-AC02-76SF00515.
DA acknowledges support from the German Federal Ministry of Economics and Technology (BMWi) provided through DLR under project 50 OR 1210.
This research has made use of the NASA/IPAC Extragalactic Database (NED) which is operated by the Jet Propulsion Laboratory, California Institute of Technology, under contract with the National Aeronautics and Space Administration.

\bibliographystyle{mn2e}
\bibliography{mnras_template}

\end{document}